\documentclass[%
 reprint, superscriptaddress, showpacs,preprintnumbers, amsmath,amssymb, prl,
]{revtex4-1}

\usepackage{graphicx}% Include figure files
\usepackage{dcolumn}% Align table columns on decimal point
\usepackage{lineno}
\usepackage{bm}% bold math
\usepackage{color}
\usepackage{txfonts}

\begin{document}

\title{\Large Topological defect launches 3D mound in the active nematic sheet of neural progenitors
}%c

\if0
\maketitle
\author{Kyogo Kawaguchi$^{1,2 *}$, Ryoichiro Kageyama$^3$ \& Masaki Sano$^{1*}$}

\begin{affiliations}
 \item Department of Physics, The University of Tokyo, Bunkyo-ku, Tokyo, 113-0033, Japan.
 \item Department of Systems Biology, Harvard Medical School, Boston, Massachusetts, 02115, USA.
 \item Institute for Virus Research, Kyoto University Sakyo-ku, Kyoto 606-8502, Japan.
\end{affiliations}

\fi

\author{Kyogo Kawaguchi$^*$}
%\email{Current email address: Kyogo\_Kawaguchi@hms.harvard.edu}
\affiliation{Department of Physics, The University of Tokyo, Bunkyo-ku, Tokyo, 113-0033, Japan.}%
\affiliation{Department of Systems Biology, Harvard Medical School, Boston, Massachusetts, 02115, USA.}
\author{Ryoichiro Kageyama}%
\affiliation{Institute for Virus Research, Kyoto University Sakyo-ku, Kyoto 606-8502, Japan.}%
\author{Masaki Sano$^*$}%
\affiliation{Department of Physics, The University of Tokyo, Bunkyo-ku, Tokyo, 113-0033, Japan.}%
\date{\today}

\maketitle

%The text is limited to 1,500 words, excluding the introductory paragraph, Methods, references and figure legends. Letters should have no more than 3-5 display items (figures and/or tables). References are limited to 30.

%This format begins with a title of, at most, 15 words, followed by an introductory paragraph (not abstract) of approximately 150 words, summarizing the background, rationale, main results (introduced by "Here we show" or some equivalent phrase) and implications of the study. This paragraph should be referenced, as in Nature style, and should be considered part of the main text, so that any subsequent introductory material avoids too much redundancy with the introductory paragraph.

\noindent
{\bf Cultured stem cells have become a standard platform not only for regenerative medicine and developmental biology but also for biophysical studies\cite{engler2006matrix,Saha2008substrate}.
Yet, the characterization of cultured stem cells at the level of morphology and macroscopic patterns resulting from cell-to-cell interactions remain largely qualitative, even though they are the simplest features observed in everyday experiments.
Here we report that neural progenitor cells (NPCs),  which are multipotent stem cells that give rise to cells in the central nervous system\cite{gotz2005cell}, rapidly glide and stochastically reverse its velocity while locally aligning with neighboring cells, thus showing features of an active nematic system\cite{Ramaswamy2003,Chate2006,Marchetti2013}.
Within the two-dimensional nematic pattern, we find interspaced topological defects with +1/2 and -1/2 charges.
Remarkably, we identified rapid cell accumulation leading to three-dimensional mounds at the +1/2 topological defects.
Single-cell level imaging around the defects allowed quantification of the evolving cell density, clarifying that not only cells concentrate at +1/2 defects, but also escape from -1/2 defects.
We propose the mechanism of instability around the defects as the interplay between the anisotropic friction and the active force field, thus addressing a novel universal mechanism for local cell density control.
}

\begin{figure*}[!hbt]
 \begin{center}
  \includegraphics[width=180mm]{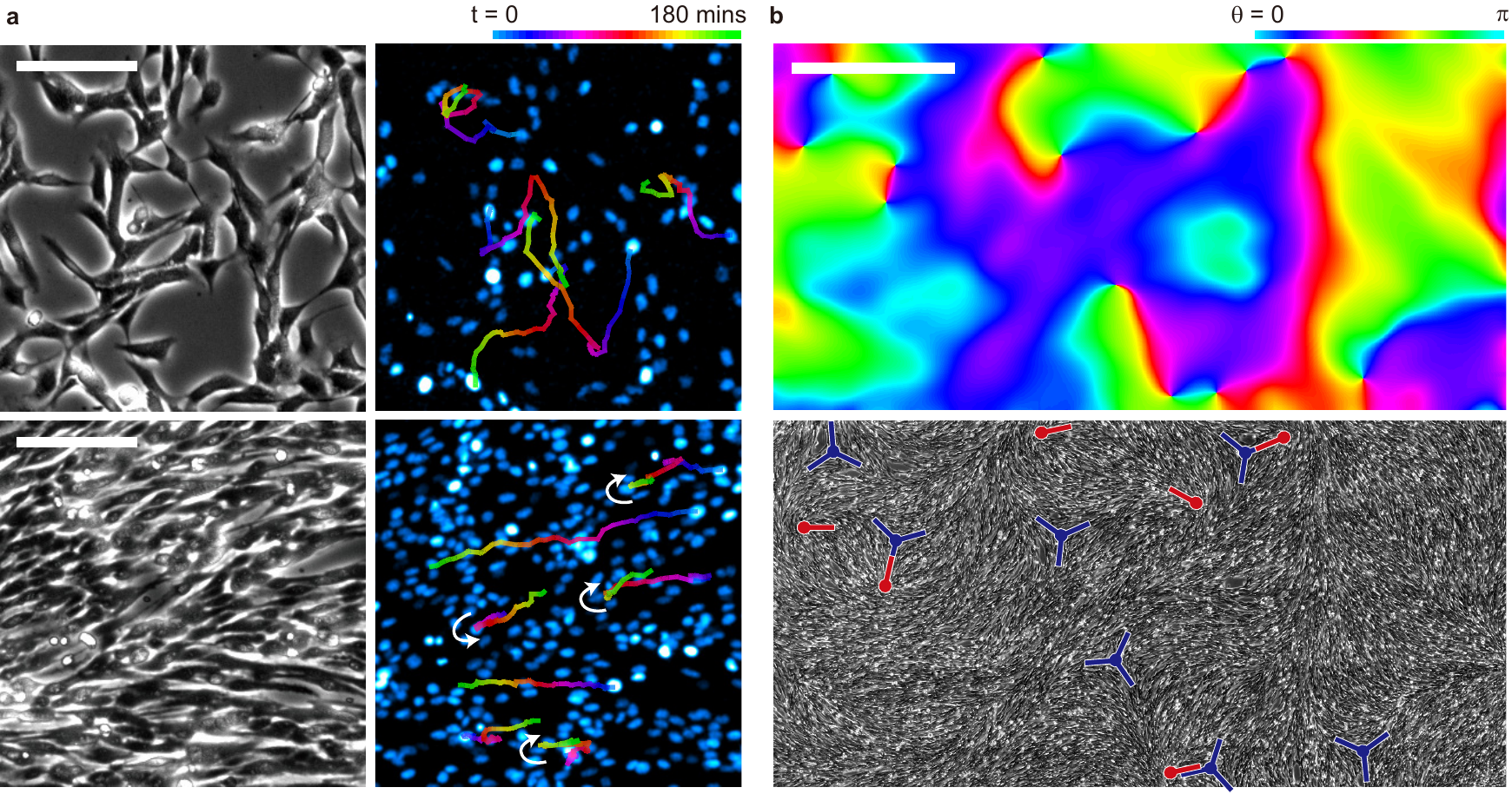}
 \end{center}
  \caption{\label{Fig1}{\bf Two-dimensional patterning and collective motion of NPCs.}
{\bf a,} Single-cell level observation of the two-dimensional culture. Top: cells in low density (1100 cells/mm$^2$) observed in phase contrast (left) and fluorescent channel  (right,  nucleus marker in pseudo-colour). Lines drawn in the right figure are the typical trajectories of single cells. Bottom: cells in high density (3000 cells/mm$^2$). The white arrows indicate velocity reversal of the tracked cell. Scale bar = 100 $\mu$m.
{\bf b,} Large field image of NPC culture. colour indicates the angle of the local alignment calculated using the tensor method (see Supplementary Information) from the phase contrast image (bottom). Singular points where all colours mix in the top figure are the topological defects, as shown explicitly in the bottom image with the charges +1/2 (red) and -1/2 (blue). Scale bar = 1~mm.
}
\end{figure*}

We used primary NPCs dissociated from fetal mice, which can survive and proliferate (i.e., undergo cell division) for numerous cell cycles under a chemically defined condition in two-dimensions\cite{conti2005niche}.
In order to elucidate the properties of cell motion and cell-to-cell interactions, we obtained live widefield images of the NPC culture.
First we found that at low density, cells tended to move in an amoeba-like fashion, therefore with no apparent order but with occasional attachment between cells (Fig.~\ref{Fig1}a).
In the high density regime, on the other hand, cells aligned with each other according to their elongated shape (Fig.~\ref{Fig1}a), which led to a macroscopic nematic liquid crystal-like pattern including topological defects (Fig.~1b, Supplementary Movie 1).
Topological defects are singular points where the orientation or alignment cannot be defined.
+1/2 and -1/2 defects are precisely a signature of underlying nematic order; they cannot exist
in polar systems, where asters and vortices will appear instead\cite{kruse2004asters}.
The alignment pattern evolved in time following the growth of the cell density (Fig.~2a), which was captured by quantifying the nematic order parameter at a fixed length scale (Supplementary Fig.~S1).

To obtain the trajectory of single cells, we established a cell line which stably expresses a fluorescent nucleus marker, H2B-mCherry.
The single cell tracking showed that local alignments at the high density regime  involved rapid motion ($\sim$40 $\mu$m/sec, Supplementary Fig.~S2) mostly in the locally aligned direction whereas in the lower density stage the motions looked more random (Fig.~2b).
From the live images it is clear that NPCs are able to slide by each other even in the seemingly densely packed configuration (Supplementary Movie 2), indicating that the interaction between the directions of cell velocity  is close to ideally nematic. 

We quantified the timescale involved in the cell motion by calculating the autocorrelation functions from the single cell trajectories (see Supplementary Information).
The autocorrelation of the direction of motion at high density fitted well with an exponential, which is a feature of a memoryless process (Fig.~2c, time frames C and D).
The switching of velocity occurred with the typical timescale of 3 hours in the high density culture  (Fig.~2c inset), and in a significantly shorter time in the lower density time frames.
To further gain insight into the property of the motion, we calculated the nematic autocorrelation function of the direction of motion (Fig.~2d).
In stark contrast to the autocorrelation, the nematic autocorrelation at high density approached zero very slowly with respect to the time interval. This indicates that the cells are changing their direction of motion 180$^\circ$ rather than randomly walking in 2D, as also illustrated in the distribution of angle relative to the longitudinal direction of motion of each cell (Supplementary Fig.~S2).

In real development, neural progenitors form epithelial layers\cite{fujita2003discovery}, where elongated cells are packed in a columnar fashion.
Within the densely packed situation, neural progenitors rapidly move their nuclei up and down within its thin cell body according to the cell cycle phase, which is known as the interkinetic nuclear migration\cite{fujita2003discovery}.
The interaction between the NPCs we observe is similar to this in respect to the smooth relative motion between neighboring cells.
To further see if the observed velocity switching is related to cell cycle, we utilized the NPC expressing the cell cycle phase monitor Fucci\cite{sakaue2008visualizing}, and found that the switching of direction occurs within the same G1 or G2 phase (Supplementary Movie 3).
This is in contrast to the interkinetic motion seen \textit{in vivo}, where the motion of nucleus is directed to the basal end (G1) or the apical end (G2) depending on its cell cycle phase\cite{fujita2003discovery}. We did, however, observe a significant difference in the time scale of switching in the two phases (Supplementary Fig.~S3), indicating that the mechanism driving the motion depend on the cell cycle phase\cite{Kosodo2011regulation}.

%Recent advances in imaging and culturing methods have brought new insights into how macroscopic structures arise in biological systems, leading to the natural challenge to elucidate laws of collective behaviors in multicellular systems.
%Recent advances in the theory of active matter has given guidelines to find characterestic features of nonequilibrium phases from the general perspective, including giant number fluctuations coupling of order and density fluctuations, and the significance of agent interactions\cite{suzuki2015polar}.
%, based on simplified microscopic models\cite{Vicsek1995,Chate2006}, active gels\cite{Prost2015}, and purely phenomenolgical arguments \cite{AditiSimha2002}.
%Although studies utilizing simple physical components and numerical simulations have brought insight,

From the general perspective of macroscopic dynamics, active nematic systems are predicted to have spatio-temporal correlations in the density and orders that are distinct from equilibrium systems\cite{Ramaswamy2003,Chate2006,Marchetti2013}.
We observed giant number fluctuation and quasi-long range order  (Supplementary Fig.~S4) in the high density NPC culture in the length scales larger than a single cell but smaller than the distance between defects. 
To see the property of correlations more carefully, we picked out areas of this intermediate size that do not contain defects, and calculated the spatial  correlations  of cell density and alignment fluctuations within the areas (Supplementary Fig.~S5).
We observed that the fluctuation of density and alignment were correlated (see Supplementary Information); the correlations were positive in the first and third quadrant, indicating that the system is an extensile active nematic system\cite{Ramaswamy2003,Marchetti2013,Edwards2009}. 
We also found that the correlation in the cell density showed an anisotropic pattern, where the typical length scale of correlation in the direction parallel to the alignment was larger than that in the direction perpendicular to it.

A clear signature of ``activeness'' in a macroscopic system can be found at the topological defects.
%, which are points in space such as asters and vortices\cite{kruse2004asters} where the direction of orientation or alignment cannot be defined.
It has been observed both in experiment and simulation\cite{Narayan2007,Kramer2003,Schaller2013,Shi2013,Giomi2013,Keber2014,DeCamp2015} that +1/2 defects typically undergo spontaneous motion in active nematic systems, which is in great contrast to defects in equilibrium liquid crystals\cite{deGennesProstBook}. 
Topological defects in the biological context have been observed in whirled grain in wood\cite{Kramer2003} as well as in tube formations in the embryonic development\cite{gilbert2013developmental}.
Defects may also yield nonequilibrium collective behavior, with the examples ranging from \textit{in vitro}\cite{Sumino2012,Keber2014,DeCamp2015} and \textit{in vivo}\cite{taniguchi2013phase} cytoskeleton dynamics to the cAMP patterning of aggregating social amoeba\cite{sawai2005autoregulatory}.

As have been observed in other two-dimensional cell cultures\cite{Duclos2014}, we clearly observed +1/2 and -1/2 defects in the dense NPC culture (Fig.~3a).
Consistent with previous non-cell studies, we occasionally observed motion of defects including the merging of defects with opposite charges (Fig.~3b).
In Fig.~3c, we show the local nematic order around the topological defects calculated from the displacement of cells inside small regions. Figure 3d shows the velocity field calculated from the same data, indicating that there is a flow of cells pointing in the direction of the front of the comet shape in +1/2 defects, whereas no clear directionality exist in the -1/2 defects.

Furthermore, the cells around the +1/2 topological defects tended to rapidly accumulate, which ended up in a three-dimensional mound of cells growing out of the observed plane (Fig.~4a, Supplementary Movie 4). 
The time dependence of cell density around the defects is shown in Fig.~4b.
To focus on the spatial fluctuation with respect to the globally growing average density, the cell density quantified from the nucleus reporter intensity was first normalized by the spatial average at a given time point. 
The radius-dependent average cell density, $\rho(R,t)$, was calculated within different diameters $R$ surrounding the defects using this normalized density (see Supplementary Information for detail).
Strikingly, we found not only that accumulation is occurring at the  +1/2 defects, but also that the density around the -1/2 defects are relatively decreasing (Fig.~4a, Supplementary Movie 5). 
The relative growth and escape rates of density near the defect centre (Fig.~4c) were comparable or larger than the global cell growth rate (0.03-0.035 per hour, Fig.~2a), which suggests that local fluctuations in cell growth will be too small to account for the density evolution, and there should be a significant net cell flow into and out from the +1/2 and -1/2 defects, respectively.

To  understand the mechanism of topological-charge dependent dynamics around the defects, we consider the theoretical setup where the local alignment close to a topological defect is given.
Using the Q-tensor representation for nematic order, the linearized time-evolution equation of the over-damped velocity field\cite{Woodhouse2013} can be written as
\begin{eqnarray}
\mbox{\boldmath $\gamma v$} = -\zeta \nabla \cdot \mbox{\boldmath $Q$}.
\end{eqnarray}
Here, the velocity field of the cells \mbox{\boldmath $v$} and the $2 \times 2$ nematic order  tensor \mbox{\boldmath $Q$} are defined locally as functions of space. The right hand side of the equation is the active force\cite{AditiSimha2002} which is the lowest order component of force that cannot derive from equilibrium free energy. The activity parameter $\zeta$ defines the nonequilibriumness of the system\cite{Edwards2009}, which is taken to be positive for the consistency with the correlation in the density-alignment fluctuation (Supplementary Fig.~S5) and the direction of the velocity field around the defect (Fig.~3c).
The friction {\boldmath $\gamma $} that is experienced by the collective cell flow originates from the stochastic switching of velocity, cell-cell interation, and the adhesive interaction between the glass substrate and the cells.
The key assumption is that {\boldmath $\gamma $} is also a locally defined tensor that depends on the alignment; it includes a term proportional to \mbox{\boldmath $Q$} in the leading order. This is reasonable given that the spatial correlation of the density is anisotropic (Supplementary Fig.~S5).

We assume Q-tensors corresponding to the alignment field around $\pm 1/2$ topological defects, with local nematic order depending only on the distance from the defect core.
Starting from a uniform density at time $t=t_0$, Eq.~(1) predicts that the density should grow as 
\begin{eqnarray}
\rho_{\pm} (R,t) \propto \exp \left[ \frac{\pm D(t-t_0)}{R^2} \right],
\end{eqnarray}
at large $R$. Here, $D$ is a positive constant that vanishes when the anisotropy of friction or the activity parameter is zero (see Supplementary Information).
Equation (2) indicates that the growth and decrease of cell density at the topological defects have the same rates with opposite sign in the plus and minus defects, which is in qualitative agreement with the observations (Fig.~4c).
Using the radial dependence of nematic order obtained by fitting the data (Supplementary Fig.~S7), and assuming the amplitude of anisotropy in friction,  we solved Eq.~(1) to obtain theoretical curves , which again showed qualitative agreement with experiment (Supplementary Fig.~S8).
The apparent difference in the $R$-dependence between theory and experiment can be due to diffusion or the motion of the nematic order, which were neglected here the simplified theory.
Nevertheless, Eq.~(1) captures the essential mechanism of how the anisotropic friction naturally lets active matter to cause density abnormality around topological defects.
  
The observed active nematic feature should be universal among other cells with elongated shape.
To see if there are other cell types that exhibit the abnormal cell density evolution at defects, we performed the same analysis in the two-dimensional culture of myoblast cells (C2C12), which also exhibited a similar nematic order.
As was expected from the final image, we found no evidence of relative density growth or decrease around the defects(Supplementary Fig.~S9). 
A typical feature of cells is that the velocity and division rates drop to almost zero in the densely packed situation\cite{Duclos2014}, a phenomena known as contact inhibition\cite{Abercrombie1954, puliafito2012collective}.
The NPC culture has an advantage that the effect of contact inhibition is limited, as well as the friction in the aligned direction being low, allowing cells to slide by its neighbors.
Remarkably, similar alignment patterns with inhomogeneous density has been seen in malign fibroblast cultures\cite{Abercrombie1979contact}, where defective contact inhibition has been associated with their malignant invasiveness.

A challenge in biophysics is to understand the laws behind the seemingly extraordinary collective dynamics observed \textit{in vivo}.
Since multicellular systems are typically out of equilibrium, the observed dynamics are typically unexplainable by clean principles such as free energy minimization.
Here we observed and analyzed the active nematic feature of the NPC culture, and found abnormal cell density evolution at +1/2 and -1/2 topological defects.
We proposed a theoretical mechanism to explain this phenomena, indicating that similar situations may universally arise when active force coexists with anisotropic friction.
Our findings clearly show how topological defects can play significant roles in collective cell dynamics, and demonstrate how the phenomenological descriptions of active matter can bring simple understandings on purely physical consequences in multicellular systems.

\noindent
{\bf Acknowledgements} 

We thank H.~Chat\'{e}, A.~M.~Klein, H.~Kori, K.~Nagai, A.~Isomura, K.~Takeuchi, D.~Nishiguchi, T.~Yamamoto, and members of R.~Kageyama lab and M.~Sano lab for fruitful discussions. We acknowledge WPI-iCeMS, Kyoto University for technical help with flowcytometry. This work was supported by Core Research for Evolutional Science and Technology (K.~K., R.~K.) and by KAKENHI (No. 25103004, "Fluctuation \& Structure") from MEXT, Japan. K.~K. acknowledges the Grant-in-Aid for JSPS Fellows (24-8031).

\noindent
{\bf Author contributions}

K.K., R.K., and M.S. designed experiments, and wrote the manuscript. K.K. performed experiments and analyses.

%\bibliography{2015kawaguchi}

\newpage

\noindent
{\bf Methods}

Two-dimensional neural progenitor cell (NPC) culture was prepared and passaged following the previously described protocols\cite{conti2005niche,Imayoshi2013} with a few modifications. Cells dissociated from primary E14 ICR mice using papain (Worthington) were cultured using DMEM/F12 (Invitrogen) supplemented with bFGF (human, 20 ng/ml, Wako), EGF (mouse,  20 ng/ml, Invitrogen), and N2 plus supplement (R\&D). Dishes were coated with  Laminin (Wako). Accutase (ICT) was used instead of papain for regular passaging. Myoblast cell line C2C12 with H2B-mCherry marker (a kind gift from A. Isomura) was cultured in DMEM (Wako) 10\% serum.

Stable cell lines of NPCs expressing nucleus markers (H2B-mcherry under PGK promoter) were established using the NEPAGENE electroporator, by co-electroporating with plasmids carrying Tol2 transposase\cite{kawakami2007tol2} and anti-Hygromycin (all plasmids were kind gifts from A. Isomura). Electroporated cells were harvested for $\sim$1 week and selected by antibiotics or sorted with BD Aria II.

For long timelapse experiments, NPCs were plated onto Laminin coated glass base dishes (Iwaki) the day before at desired cell densities. Tiled fluorescent and phase contrast imaging was performed by AF6000 and AF7000 (Leica Microsystems) epifluorescent microscopes.  C2C12 was similarly prepared but without the Laminin coating.

Colour maps of phase on the cell alignment images and the automatic detection of topological defect points were obtained by a MATLAB script implementing the tensor method\cite{jahne1993spatio,Rezakhaniha2012}. Briefly, the gradient tensor of the intensity matrix was calculated for the raw phase contrast images, and the  principle angles and nematic orders were calcuated from these tensors. See Supplementary Information for further detail.

To count the cells in the fluorescent images for Fig.~2a,S1, and S4, the raw images were first median filtered (2~pixels = 1.8~$\mu$m). After suppressing small maxima (\texttt{imhmax} in MATLAB with a manually set threshold), the number of regional maxima (\texttt{imregionalmax} in MATLAB) was counted.

For the tracking of cells for Figs.~1a,2b,c,d,3c,d,S1,S5 and S6, we developed a nucleus marker tracker (MATLAB script). The position of a cell was defined by the centroid of the cell nucleus fluorescent signal, and was traced in the subsequent time frames based on a cost function of position and the fluorescent intensity.   Lost cells or cells tracked but with high value in cost function were excluded from the data analyses, as can be seen in the Supplementary Movies 2 and 3. To obtain long and precise single cell tracking data for Figs.~2b,c,d, we further added a manual option in our algorithm to detect failure of tracking and hault to let the operator manually point to the correct position of the cell. All MATLAB codes are accessible upon request.

\newpage

\renewcommand{\theequation}{S\arabic{equation}}
\setcounter{equation}{0}

\section{Image analysis}

Here we present the methods of image analysis we used to generate the main and supplementary figures, and provide definitions of plotted and discussed quantities.

\subsection{Tensor method and fluctuation of alignment}

In order to quantify the local alignment of the cells, we used the tensor method which is a powerful tool to analyze two-dimensional patterns\cite{jahne1993spatio,Rezakhaniha2012}. For an acquired raw phase contrast image matrix $I({\bf r},t)$, where ${\bf r}= (x,y) $ and $t$ denote the two-dimensional position (pixel) and time (frame number), respectively, we directly calculated the differential tensor:
\begin{eqnarray}
I'({\bf r},t) :=  \left( \begin{array}{cc}
(\Delta_x I)^2 & \Delta_x I\Delta_y I  \\
\Delta_x I\Delta_y I & (\Delta_y I)^2 \end{array} \right).
\end{eqnarray}
Here,
\begin{eqnarray}
\Delta_x I  &:=& I(x+1,y)-I(x-1,y) \\
\Delta_y I &:=& I(x,y+1)-I(x,y-1).
\end{eqnarray}
We omitted the notation for the position and time dependence for simplicity.
After applying a spatial Gaussian filter to $I'$ to obtain
\begin{eqnarray}
\widetilde{I}'({\bf r},t) :=  \left( \begin{array}{cc}
\widetilde{I}'_{xx}  & \widetilde{I}'_{xy}   \\
\widetilde{I}'_{xy}  & \widetilde{I}'_{yy}  \end{array} \right),
\end{eqnarray}
where the size of the filter was typically 3$\sim$5 cells in length, we calculated the local phase (angle of alignment)
\begin{eqnarray}
\theta ({\bf r},t) &:=& \frac{1}{2}\arctan \left( \frac{ 2 \widetilde{I}'_{xy}}{\widetilde{I}'_{xx}-\widetilde{I}'_{yy}} \right ).
\end{eqnarray}
The color plots in Figs.~1b,3a,b and the position of the topological defects used in the analysis for Figs.~4b,c,S9  were obtained from $\theta ({\bf r},t)$.
%We also obtained the coherence
%\begin{eqnarray}
%{\rm Coh}({\bf r},t)&:=& \frac{(\widetilde{I}'_{xx}-\widetilde{I}'_{yy})^2+4 \widetilde{I}'^2_{xy} }{ ( \widetilde{I}'_{xx} + \widetilde{I}'_{yy} )^2},
%\end{eqnarray}
%which we compared with the local order parameter (see following section).

In a manually chosen small region $C_d$ where cells are aligned without a topological defect, the global direction of alignment was defined as
\begin{eqnarray}
\bar{\theta}(t) &:=& \frac{1}{2}\arctan \left( \frac{ 2 \bar{I}'_{xy}}{\bar{I}'_{xx}-\bar{I}'_{yy}} \right ).
\end{eqnarray}
where $\bar{I}'_{xx}, \bar{I}'_{xy},$ and $\bar{I}'_{yy}$ are components of the averaged differential tensor:
\begin{eqnarray}
\bar{I}'(t):= \frac{1}{N_{d}} \sum_{{\bf r} \in D} \widetilde{I}'({\bf r},t) ,
\end{eqnarray}
where $N_d$ is the number of pixels in the region $C_d$.
The local fluctuation of alignment was then defined as
\begin{eqnarray}
\delta n ({\bf r},t) := \frac{1}{2}\sin \left\{ 2 \left[\theta({\bf r},t) - \bar{\theta} \right] \right\} \label{eq:flucn}
\end{eqnarray}

\subsection{Normalized cell density from fluorescent signal}
The fluorescent channel image for the nucleus marker was put through a 2 pixel (1.8~$\mu$m) median filter to obtain the image matrix $F({\bf r}, t)$. The background of fluorescent signal $B(t)$ was calculated as the average value of the bottom 0.15 fraction of the total pixels, which is an empirically chosen fraction. Assuming that the fluctuation in the signal of fluorescent intensity comes from a linear effect of the light source fluctuation, the intensity of fluorescence was standardized as
\begin{eqnarray}
\widetilde{F} ({\bf r}, t):= \frac{F({\bf r}, t)-B(t)}{B(t)}.
\end{eqnarray}
For the thin line in Fig.~2a, we calculated the spatial average of $\widetilde{F} ({\bf r}, t)$ obtained in a widefield (2.3~mm$\times$2.3~mm)  as a function of time in a 24 hour timelapse movie, and multiplied a constant (1.1$\times 10^{-3}$) to let it match with the cell density obtained by cell counting in the time frame C.

The intensity fluctuation plotted in Fig.~S2 was obtained from $\widetilde{F} ({\bf r}, t)$. Denoting the $m$-th square region of size $L$ as $D_m(L)$, the intensity inside the region is simply
\begin{eqnarray}
I_m (L, t) := \sum_{{\bf r} \in D_m(L)} \widetilde{F} ({\bf r}, t).
\end{eqnarray}
The average and fluctuation of $I_m (L, t)$ was calculated by
\begin{eqnarray}
 I (L, t)  &:=& \frac{1}{N_{L}} \sum_{m} I_m (L, t) \\
\Delta I (L, t)  &:=& \sqrt{\frac{1}{N_{L}-1} \sum_{m} \left [ I_m (L, t) - I (L, t)  \right] ^2},
\end{eqnarray}
where $N_L$ is the number of non-overlapping square regions ($2^6 \sim 2^{14}$) defined within the large image.

To further take into account the trend of growth of the fluorescent signal due to cell proliferation, we introduced the normalized density
\begin{eqnarray}
\rho({\bf r}, t):= \frac{\widetilde{F} ({\bf r}, t)} {\sum_{\bf r} \widetilde{F} ({\bf r}, t)/N_A},
\end{eqnarray}
where $N_A$ is the number of pixels in the area of sum. The spatial fluctuation of cell density was then obtained by
\begin{eqnarray}
\delta \rho({\bf r}, t) := \rho({\bf r}, t) -1. \label{eq:flucrho}
\end{eqnarray}

For the analysis of cell density around the topological defects, we calculated the normalized cell density for regions with different sizes:
\begin{eqnarray}
 \rho _{\pm }(R, t) := \frac{1}{N_{C_{\pm}(R)}}\sum_{{\bf r} \in C_{\pm}(R)} \rho({\bf r}, t) , \label{eq:flucrho}
\end{eqnarray}
where $C_{\pm}(R)$ is the circular region with diameter $R$ around the +1/2 or -1/2 topological defect with the number of pixels $N_{C_{\pm}(R)}$. The cell densities $\rho _{\pm }(R, t) $ were averaged over 4 (+1/2) and 5 (-1/2) defects and plotted in Fig.~4b,c.

\subsection{Spatial correlation of cell density and alignment}
Using the fluctuations calculated by Eqs.~(\ref{eq:flucn},\ref{eq:flucrho}), we obtained the spatial correlation functions
\begin{eqnarray}
C _{\rho \rho}({\bf r}) &:=& \frac{\langle \delta \rho({\bf r'}, t) \delta \rho({\bf r'} + {\bf r}, t) \rangle}{\langle \delta \rho({\bf r'}, t)^2 \rangle} \label{eq:corrrhorho}\\
C _{\rho n}({\bf r}) &:=& \frac{\langle \delta \rho({\bf r'}, t') \delta n ({\bf r'} + {\bf r}, t) \rangle}{\sqrt{\langle \delta \rho({\bf r'}, t)^2 \rangle \langle \delta n({\bf r'}, t)^2 \rangle}} \label{eq:corrrhon}\\
C _{n n}({\bf r}) &:=& \frac{\langle \delta n({\bf r'}, t) \delta n({\bf r'} + {\bf r}, t) \rangle}{\langle \delta n({\bf r'}, t)^2 \rangle}, \label{eq:corrnn}
\end{eqnarray}
where
\begin{eqnarray}
\langle \delta \rho({\bf r'}, t) \delta \rho({\bf r'} + {\bf r}, t) \rangle := \frac{1}{T}\sum_t \frac{1}{N_{\bf r}}\sum _{\bf r'}\delta \rho({\bf r'}, t) \delta \rho({\bf r'} + {\bf r}, t),
\end{eqnarray}
etc. Here, $N_{\bf r'}$ is the number of contributions in the sum, and $T$ is the number of time frames used in the analysis. The spatial correlations were averaged over  4 different small regions and plotted in Fig.~S5.

%We also obtained the temporal correlations of the fluctuations in a similar manner:
%\begin{eqnarray}
%C _{\rho \rho}(\tau) &:=& \frac{\langle \delta \rho({\bf r'}, t) \delta \rho({\bf r'} , t + \tau) \rangle}{\langle \delta \rho({\bf r'}, t)^2 \rangle} \label{eq:corrTrhorho}\\
%C _{\rho n}(\tau) &:=& \frac{\langle \delta \rho({\bf r'}, t) \delta n ({\bf r'} , t+\tau) \rangle}{\sqrt{\langle \delta \rho({\bf r'}, t)^2 \rangle \langle \delta n({\bf r'}, t)^2 \rangle}} \label{eq:corrT+rhon}\\
%C _{n n}(\tau) &:=& \frac{\langle \delta n({\bf r'}, t) \delta n({\bf r'} , t+\tau) \rangle}{\langle \delta n({\bf r'}, t)^2 \rangle}. \label{eq:corrTnn}
%\end{eqnarray}
%The results are plotted in Fig.~S4.

\subsection{Cell tracking for nematic order and autocorrelation calculations}

In order to elucidate the single cell level features in the nematic patterning, we implemented two cell tracking methods.

In the first method, the displacement  ${\bf d}_j (t)$ of the $j$-th cell at position ${\bf r}_j(t)$ was obtained by connecting the nearest regional max points in the fluorescent image between subsequent time frames (5 min interval):
\begin{eqnarray}
{\bf d}_j (t) := {\bf r}_j(t+1) - {\bf r}_j(t) =  d_j(t)\left( \begin{array}{cc}
 \cos \theta_j (t)  \\
\sin \theta_j (t) \end{array} \right).
\end{eqnarray}
where $\theta_j (t)$ is the instantaneous velocity angle.
Average velocity as a function of time plotted in Fig.~S2 was obtained by taking the average of the absolute values of displacements.
In Fig.~S6, we show an example of ${\bf d}_j (t)$ obtained in a large field of view. In case where $d_i(t)=0$ (no cell motion in the resolution of a pixel), $\theta_j (t)$ was replaced with a random variable sampled from $[0,2 \pi)$, although its effect was negligible. This simple method led to a $\sim$10\% error of connecting wrong cells in the sequences, although after averaging over finite sized boxes, the obtained pattern was very similar to the one obtained by the tensor method from a static phase contrast image (Fig.~S6).

Again by introducing the $m$-th box of size $L$ as $D_m(L)$,  the global nematic order was calculated by
\begin{eqnarray}
S(L,t) := \left \langle \cos 2\theta_j (t) \right \rangle_L ^2 + \left\langle \sin 2\theta_j (t) \right\rangle_L ^2, \label{nematicorder}
\end{eqnarray}
with
\begin{eqnarray}
\left\langle \cos 2\theta_j (t) \right\rangle_L &:=& \frac{1}{N_L} \sum_m \frac{1}{N_m} \sum_{j,{\bf r}_j \in D_m(L)} \cos 2\theta_j (t) \\
\left\langle \sin 2\theta_j (t) \right\rangle_L &:=& \frac{1}{N_L} \sum_m \frac{1}{N_m} \sum_{i,{\bf r}_j \in D_m(L)} \sin 2\theta_j (t).
\end{eqnarray}
where $N_m$ is the number of cells in the $m$-th box.
$S(L,t)$ as a function of $L$ for various $t$ is shown in Fig.~2e.

The second method involves tracking fewer single cells for a longer period of time. We similarly obtained the displacement of the $j$-th cell ${\bf d}_j(t)$ and its position ${\bf r}_j(t)$, but for sequential time frames keeping track of the same cell. See Supplementary movie 1 and Fig.~1a.

From the time series of displacements, the cosine-autocorrelation of the angle was obtained as
\begin{eqnarray}
C^{\rm pol}(\tau) &:=& \langle \cos [\theta_j (t+\tau) -\theta_j (t)] \rangle \nonumber \\
&=& \frac{1}{N} \sum_j \frac{1}{T_j-\tau}\sum_{t} \cos [\theta_j (t+\tau) -\theta_j (t)],
\end{eqnarray}
where $T_i$ is the number of frames $-1$ in the trajectory of the $j-$th cell and $N$ is the number of cells taken in the ensemble.
Similarly, the nematic-autocorrelation was calculated as
\begin{eqnarray}
C^{\rm nem}(\tau) &:=& \langle \cos [2\theta_j (t+\tau) -2\theta_j (t)] \rangle \nonumber \\
&=& \frac{1}{N} \sum_j \frac{1}{T_j}\sum_{t} \cos [2\theta_j (t+\tau) -2\theta_j (t)].
\end{eqnarray}
$C^{\rm pol}(\tau)$ and $C^{\rm nem}(\tau)$ for the nucleus marker tracked data were plotted in Figs.~2c and d, respectively. Similar analysis was conducted for cell cycle specified nucleus tracking data, as shown in Fig.~S4a,b.
Obtained curves were fitted with a simple exponential function $A\exp(-2\tau /\tau_{\rm r})$, where $A$ is a coefficient and $\tau_{\rm r}$ is the time scale of velocity reversal. The flipping time scale $\tau_{\rm f}$ as a function of cell density was plotted in the inset of Fig.~2c.

For each cell trajectory, the longitudinal angle $\Theta_j$ was determined by
\begin{eqnarray}
\Theta_j := \frac{1}{2} \arctan \left( \frac{\langle \sin 2\theta_j (t) \rangle_j }{\langle \cos 2\theta_j (t) \rangle_j } \right)
\end{eqnarray}
where
\begin{eqnarray}
\langle \cos 2\theta_j (t) \rangle_j &:=& \frac{1}{T_j} \sum_t \cos 2\theta_j (t) \\
\langle \sin 2\theta_j (t) \rangle_j &:=& \frac{1}{T_j} \sum_t \sin 2\theta_j (t)
\end{eqnarray}
In Fig.~S2 and S3 we plotted the distribution of $\theta_j(t) - \Theta_j$.

\newpage

\section{Theory}

Here we consider the theoretical mechanism of the cell accumulation at +1/2 defects and escape from -1/2 defects. Our aim is to present an analytical understanding of the theory and its correspondence to the experimental observation.

\subsection{Setup}
In a simplified linear theory, the deterministic time evolution of the Q-tensor field and the velocity field is phenomenologically given by an overdamped stokes equation\cite{Woodhouse2013}:
\begin{eqnarray}
\mbox{\boldmath $\gamma v$} = - \nabla \cdot \left( \zeta \mbox{\boldmath $Q$} + \alpha S \mbox{\boldmath $I$}  \right), \label{eq:fricbalance}
\end{eqnarray}
which is the balance between the friction force and the active nematic force. 
Here, $\mbox{\boldmath $I$} $ is the identity matrix, and the viscosity and nonlinear terms are neglected.
The friction term is introduced to account for the fact that the cells will eventually damp its velocity field to zero even in the case of spatially uniform or zero nematic order. The coefficient $\zeta$ controls the system to be extensile ($\zeta >0$) or contractile ($\zeta < 0$).
The active nematic stress includes a term proportional to $\alpha$, which depends on the nematic order $S$.
Note that the constant $\alpha$ is typically set as 0 assuming an incompressible fluid, but here we leave it as finite as the cell density inhomogeneity is non-negligible. Nevertheless, we will be forced to set it as zero in order to avoid finite flux at $r=0$.

We further assume that the friction {\boldmath $\gamma$} is an anisotropic tensor which depends on the local orientation order and angle. In the lowest order, it should read
\begin{eqnarray}
\mbox{\boldmath $\gamma$} = \gamma_0 \mbox{\boldmath $I$} - \Delta \mbox{\boldmath $Q$} \label{eq:gammadef}
\end{eqnarray}
where $\gamma_0$ is the orientation independent friction, and $\Delta$ is the scalar parameter representing the extent of anisotropy.
This assumption allows for example the friction to be larger when the velocity is perpendicular to the orientation angle ($\Delta>0$).

The description of dynamics is closed by introducing the equation of motion of the alignment field  \mbox{\boldmath $Q$}\cite{Marchetti2013,Ramaswamy2003}. Seeing however that the motion of the orientation field is slow, we treat it as effectively fixed for the timescale of interest.
Under this assumption, we can use Eq.~(\ref{eq:fricbalance}) to calculate the velocity field for a  given orientation field {\boldmath $Q$}.
In the vicinity of $+1/2$ and $-1/2$ topological defects located at the origin $(0,0)$, the Q-tensors should be well approximated by
\begin{eqnarray}
\mbox{\boldmath $Q$}^+ (r,\theta) &=& S_+(r)   \left(
    \begin{array}{cc}
      \cos \theta & \sin \theta   \\
      \sin \theta & -\cos \theta 
    \end{array}
  \right)  \label{eq:pdefect}\\
\mbox{\boldmath $Q$}^- (r,\theta) &=& S_-(r)  \left(
    \begin{array}{cc}
      \cos \theta & - \sin \theta   \\
      - \sin \theta & -\cos \theta 
    \end{array}
  \right) \label{eq:mdefect},
\end{eqnarray}
where we assumed that the order parameters $S(r)=S_+ (r)$ and $S_- (r)$ corresponding to plus and minus 1/2 defects, respectively, do not depend on $\theta$. Note that $S_+ (r )$ and $S_- (r )$ should vanish at $r \to 0$.

\subsection{Velocity field around topological defects}

From Eqs.~(\ref{eq:pdefect},\ref{eq:mdefect}),  we first have
\begin{eqnarray}
\nabla \cdot \mbox{\boldmath $Q$}^+ &=& \left[ S_+' (r) + \frac{S_+ (r)}{r} \right] \left(
    \begin{array}{c}
      1   \\
      0 
    \end{array}
  \right) \label{eq:pdiv} \\
\nabla \cdot \mbox{\boldmath $Q$}^- &=& \left[ S_-' (r) - \frac{S_- (r)}{r} \right] \left(
    \begin{array}{c}
      \cos 2\theta   \\
      - \sin 2 \theta 
    \end{array}
  \right).   \label{eq:mdiv} 
\end{eqnarray}
Solving Eq.~(\ref{eq:fricbalance}), we obtain the velocity fields around the plus and minus 1/2 defects:
\begin{eqnarray}
\mbox{\boldmath $v$} ^+ &=& c^+(r)
\left(
    \begin{array}{c}
      1    \\
     0
    \end{array}
  \right) +v^+ (r) 
\left(
    \begin{array}{c}
      \cos \theta    \\
     \sin \theta
    \end{array}
  \right)
\label{eq:pv} \\
\mbox{\boldmath $v$} ^- &=&  c^-(r)
\left(
    \begin{array}{c}
      \cos 2\theta    \\
    \sin 2\theta
    \end{array}
  \right) +v^-(r) 
\left(
    \begin{array}{c}
      \cos \theta    \\
     \sin \theta
    \end{array}
  \right) \label{eq:mv} 
\end{eqnarray}
where we introduced 
\begin{eqnarray}
c^\pm(r) &:= &-[  \zeta
\gamma_0 f ^\pm (r) + \alpha \Delta S_\pm(r) g ^\pm (r)] \\
v^\pm(r) &:= &-[ \zeta \Delta S_\pm(r) f ^\pm (r) + \alpha \gamma_0 g ^\pm (r)] \label{eq:tanvel} \\
f ^\pm (r) &=& 
\frac{ S_\pm' (r) \pm S_\pm (r) /r }{ \gamma_0 ^2 -  \Delta^2 S_\pm (r)^2}   \label{eq:pf} \\
g ^\pm (r) &=& 
\frac{ S_\pm' (r)  }{ \gamma_0 ^2 -  \Delta^2 S_\pm (r)^2} .  \label{eq:pg} 
\end{eqnarray}
In the limit of $r \to 0$, $v^\pm(r)$ can be finite when $\alpha$ and $S'_{\pm} (r \to 0)$ are finite, which makes the flux diverge at the defect core. Since it is typical to consider $S'_{\pm} (r \to 0) \neq 0$ for topological defects, we set $\alpha =0$ in order to obtain a physically relevant velocity field.

\subsection{Motion of defects and the radial flux of cells}

From Eqs.~(\ref{eq:pv},\ref{eq:mv}) we first find that the +1/2 defect has a net velocity but the -1/2 defect does not. To see this, consider a finite sized circle $C_R$ around the origin with radius $R$. The average velocities within the circle are
 \begin{eqnarray}
\mbox{\boldmath $v$}_{C} ^+ &:=& \frac{1}{\pi R^2} \int_{C_R} d{\bf r}  \mbox{\boldmath $v$}^+ =
\int_0 ^R dr \frac{2r  c^+ (r)}{ R^2}
\left(
    \begin{array}{c}
      1    \\
     0
    \end{array}
  \right)\\
 \mbox{\boldmath $v$}_C ^- &:=& \frac{1}{\pi R^2} \int_{C_R} d{\bf r}  \mbox{\boldmath $v$}^- = 0.
\end{eqnarray}

Secondly, the cell density evolution around the defects can be captured by introducing the tangentially averaged velocity at radius $R$:
 \begin{eqnarray}
v^\pm (R) = \frac{1}{2 \pi R} \int_0 ^{2\pi} d \theta \left(
    \begin{array}{c}
      R \cos \theta    \\
     R \sin \theta
    \end{array}
  \right)
 \cdot \left. \mbox{\boldmath $v$}^\pm \right|_{r=R} ,
\end{eqnarray}
which is equivalent to Eq.~(\ref{eq:tanvel}).
Defining the defect-centered cell density as
 \begin{eqnarray}
\rho^\pm (R,t) := \frac{1}{\pi R^2} \int_C d{\bf r} \rho ({\bf r},t),
\end{eqnarray}
the continuity equation reads
 \begin{eqnarray}
\frac{\partial}{\partial t}\rho^\pm (R,t) = - \frac{v^\pm (R)}{R^2} \frac{\partial}{\partial R} \left[ R^2 \rho^\pm (R,t) \right ]. \label{eq:cont}
\end{eqnarray}

Let us assume $S_\pm (R) = S_0(1- e^{-R/R_0})$, as fitted in Fig.~S6. Then, using $\delta := \Delta S_0 / \gamma_0$, we have
\if0
  \begin{eqnarray}
v^\pm (R) =- \frac{ S_0 [\zeta \Delta S_0 (1-e^{-R/R_0}) ( e^{-R/R_0}/R_0 \pm (1-e^{-R/R_0})/R )+ \alpha \gamma_0 e^{-R/R_0}/R_0 ] }{ \gamma_0 ^2 -  \Delta^2 S_0^2  (1-e^{-R/R_0})^2} 
\end{eqnarray}
\fi
\begin{eqnarray}
v^\pm (R) &=&- \frac{ \zeta S_0 \delta }{ \gamma_0} \frac{ 1}{ 1 -  \delta^2  (1-e^{-R/R_0})^2}  \nonumber \\
 &\times& [ (1-e^{-R/R_0}) ( e^{-R/R_0}/R_0 \pm (1-e^{-R/R_0})/R ) ]
\end{eqnarray}
In the limit of large $R$, we have
  \begin{eqnarray}
v ^+ (R \gg R_0)   \simeq - v ^- (R \gg R_0)  \simeq - \frac{ \zeta S_0 \delta}{ \gamma_0 \left( 1 -  \delta^2 \right)R },  \label{eq:dc}
\end{eqnarray}
which explains qualitatively why the defects with opposite charges have different signs in the density growth rate. The constant $D$ presented in Eq.~(2) of the main text is thus given by
  \begin{eqnarray}
D:= \frac{2 \zeta S_0 \delta}{ \gamma_0 \left( 1 -  \delta^2 \right) },  \label{eq:dc}
\end{eqnarray}

Starting with a spatially uniform cell density at $t=0$, we solved Eq.~(\ref{eq:cont}) numerically using the parameters $S_0$ and $R_0$ obtained by fitting in Fig.~S7. The result for the case of $\Delta= 0.2\gamma_0$ is given in Fig.~S8. 

To compare our theory with experiment, we estimated the $R$-dependent growth rate $g^\pm (R)$ by assuming that the time evolution of the densities $\rho^\pm (R,t)$ will roughly follow
\begin{eqnarray}
\rho^\pm (R,t) \simeq {\rho}^\pm_0(R)\exp \left[t g^\pm (R)  \right]  .  \label{eq:rhobar}
\end{eqnarray}
Here, $\bar{\rho}^\pm (R)$ is the density at $t=0$. We obtained $g^\pm (R) $ by fitting $\log \rho^\pm (R,t)$ with a first order polynomial of $t$ for each $R$ and plotted in Fig.~4c and Fig.~S8a.

Although the opposite signs of the growth rates in the $\pm$1/2 defects is well captured by the theory, the $R$-dependence seems to be sharper in the numerical result (Fig.~S8) compared with experimental curves (Fig.~4c). The actual system is obviously more complex than our simplified theoretical setup; it should include diffusion, motion of the nematic order, a detailed $\theta$-dependent structure near the defect, and the finiteness of the particle size should be taken into account. We note however that none of these excluded effects have  obvious reasons to enhance the cell density growth/decrease around the defects.

\section{Supplementary Videos}

\noindent
{\bf Supplementary Movie 1: Widefield live image of neural progenitor cell culture at high density.}
Live phase contrast (left) and fluorescent (right) images taken over 24 hours. Fluorescent signals from H2B-mCherry (nucleus marker) is shown by pseudo-color. 

\vspace{5mm}
\noindent
{\bf Supplementary Movie 2: Tracking of single cells using nucleus marker.}
White: Fluorescent live image from H2B-mcherry in the high-density culture of neural progenitors (24 hours). Green boxes are the results of the half-manual tracking of single cells.

\vspace{5mm}
\noindent
{\bf Supplementary Movie 3: Tracking of single cells using Fucci.} 
Cells at G1 phase (red) and G2 phase (green) observed by the Fucci marker through fluorescent microscopy (24 hours). The boxes represent the results of the half-manual tracking method.

\vspace{5mm}
\noindent
{\bf Supplementary Movie 4: Live image around a +1/2 topological defect.} 
Live phase contrast (left) and fluorescent (right) images around a +1/2 topological defect over 24 hours.

\vspace{5mm}
\noindent
{\bf Supplementary Movie 5: Live image around a -1/2 topological defect.} 
Live phase contrast (left) and fluorescent (right) images around a -1/2 topological defect over 24 hours.

\begin{figure*}[!ht]
 \begin{center}
  \includegraphics[width=180mm]{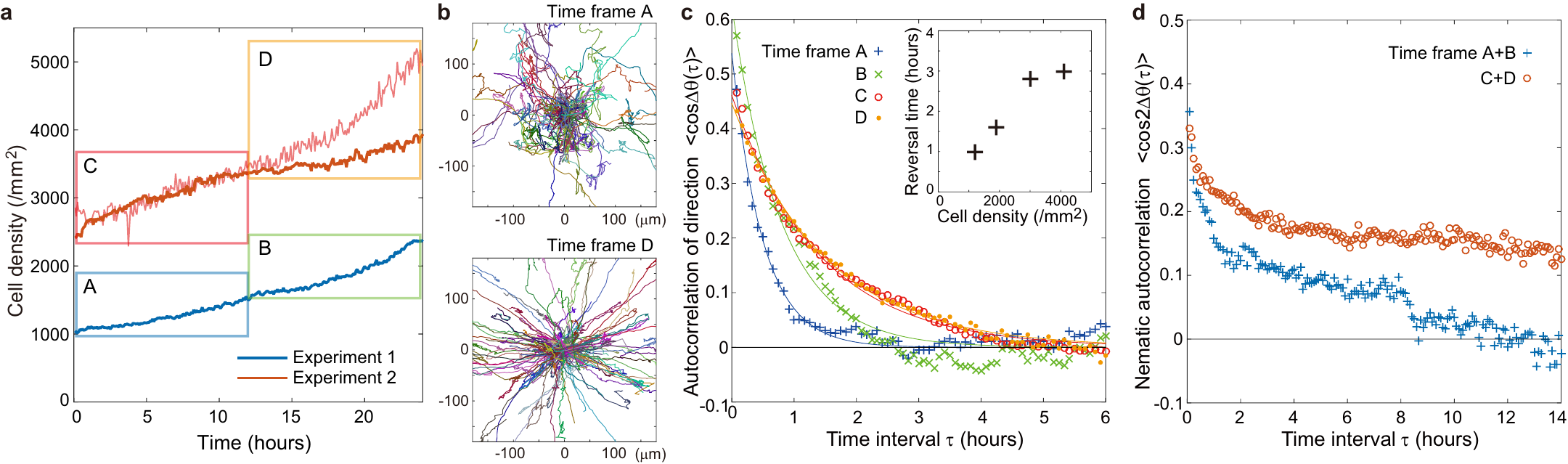}
 \end{center}
  \caption{\label{Fig2} 
{\bf Single-cell tracking reveals transition from a random to bipolar motion at high density.}
{\bf a,} Time dependence of cell density obtained from two 24 hour time-lapse movies by the counting of cell nuclei in the whole images. At high density (time frame D) the counting method fails due to cell accumulation. The thin line corresponds to the cell density estimated by integrating the fluorescent signal intensity from the nuclei in experiment 2.  
{\bf b,} Trajectories of single cells shown in different colours within the time frames A and D from {\bf a}. The starting points of each track were set at the origin in the plots.
{\bf c,} Autocorrelation of the instantaneous direction of motion generated from 61 (for time frames A and B) and 169 (for time frames C and D) randomly tracked cells as a function of time interval $\tau$. Inset shows the velocity reversal time estimated by fitting the autocorrelation data with exponential (solid lines, see Supplementary Information).
{\bf d,} Nematic autocorrelation of the instantaneous direction of motion generated from the same data as {\bf c} but with pairs of time frames combined in order to quantify the long time behavior.
}
\end{figure*}

\begin{figure*}[!t]
 \begin{center}
  \includegraphics[width=180mm]{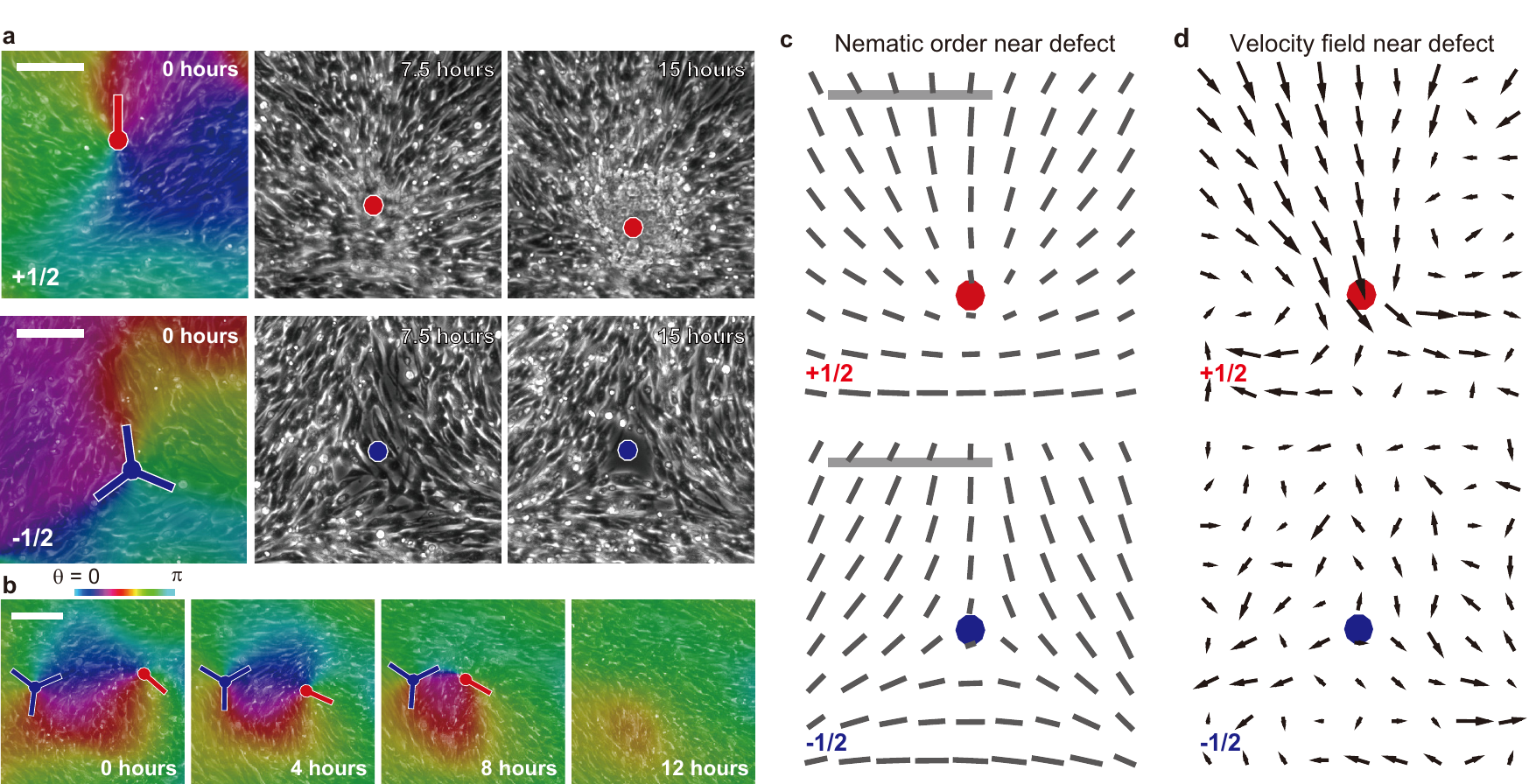}
 \end{center}
  \caption{\label{Fig3}
{\bf Active topological defects.}
{\bf a,} Typical time course of topological defects in the laboratory frame at high cell density. The red and blue spots correspond to the +1/2 and -1/2 topological defects, respectively.
{\bf b,} Merging of +1/2 and -1/2 defects resulting in annihilation.
{\bf c,} Local nematic order (length, a.u.) and alignment direction obtained by particle tracking.  Average of 4 (+1/2) and 6 (-1/2) independent defects over the time frames ($\sim$ 18 hours each) before cell accumulation becomes significant. 
{\bf d,} Local velocity field obtained from the same data. The lattice points corresponds to those in {\bf c}. The lengths of arrows are proportional to the velocity, with the same units in the top and bottom figures.
All scale bars = 100 ${\rm \mu m}$. }
\end{figure*}

\begin{figure*}[!t]
 \begin{center}
  \includegraphics[width=151mm]{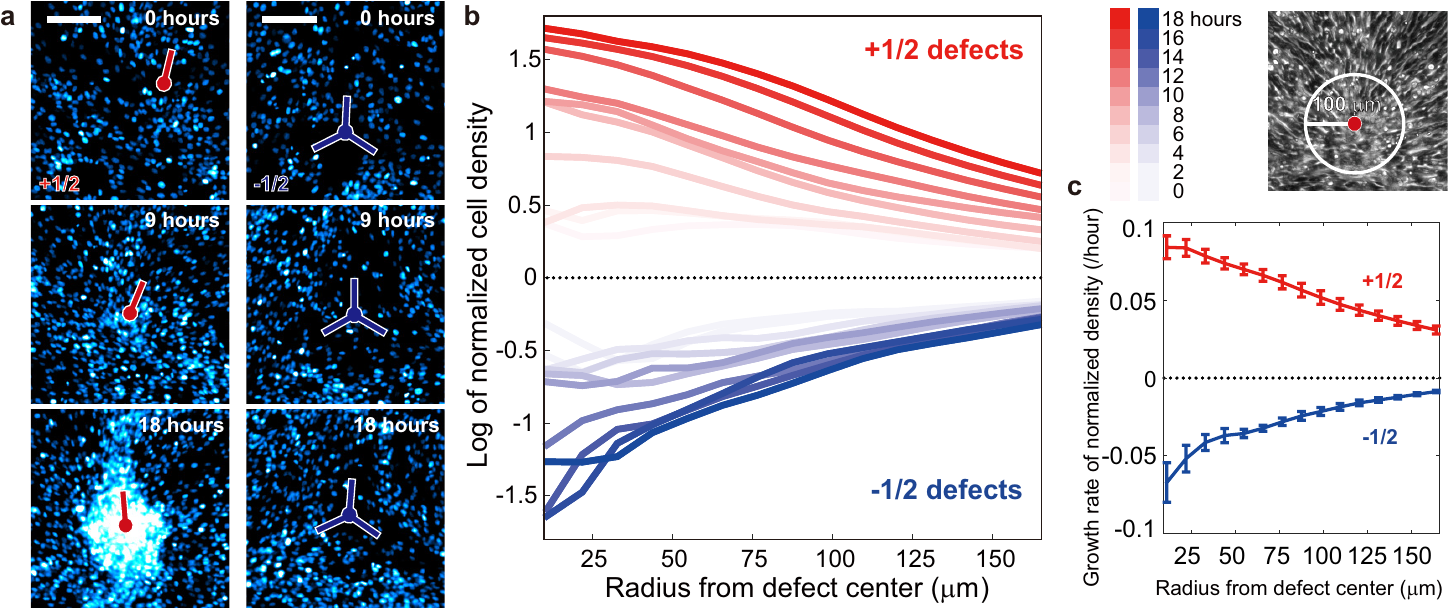}
 \end{center}
  \caption{\label{Fig4}
{\bf Cell accumulation and dispersion around the topological defects.}
{\bf a,} Time course of +1/2 (left) and -1/2 (right) topological defects in the fluorescent channel (pseudo-colour). Scale bars = 100 $\mu$m.
{\bf b,} Log of the cell density inside a circle centred around topological defects as a function of the radius of the circle, $\log [\rho_\pm (R,t)]$. Cell density was calculated from the fluorescent signal of the nuclei marker, and normalized by the time-dependent (growing) cell density after subtracting background (see Supplementary Information). Average over 4 and 5 defects for +1/2 and -1/2 defects, respectively, observed in the same dish and time course. {\bf c,} Growth rate of normalized cell density as a function of the radius. Error bars are fitting errors. See Supplementary Information for detail. 
}
\end{figure*}

\renewcommand\thefigure{S\arabic{figure}}    
\setcounter{figure}{0}

\begin{figure*}[!t]
 \begin{center}
  \includegraphics[width=130mm]{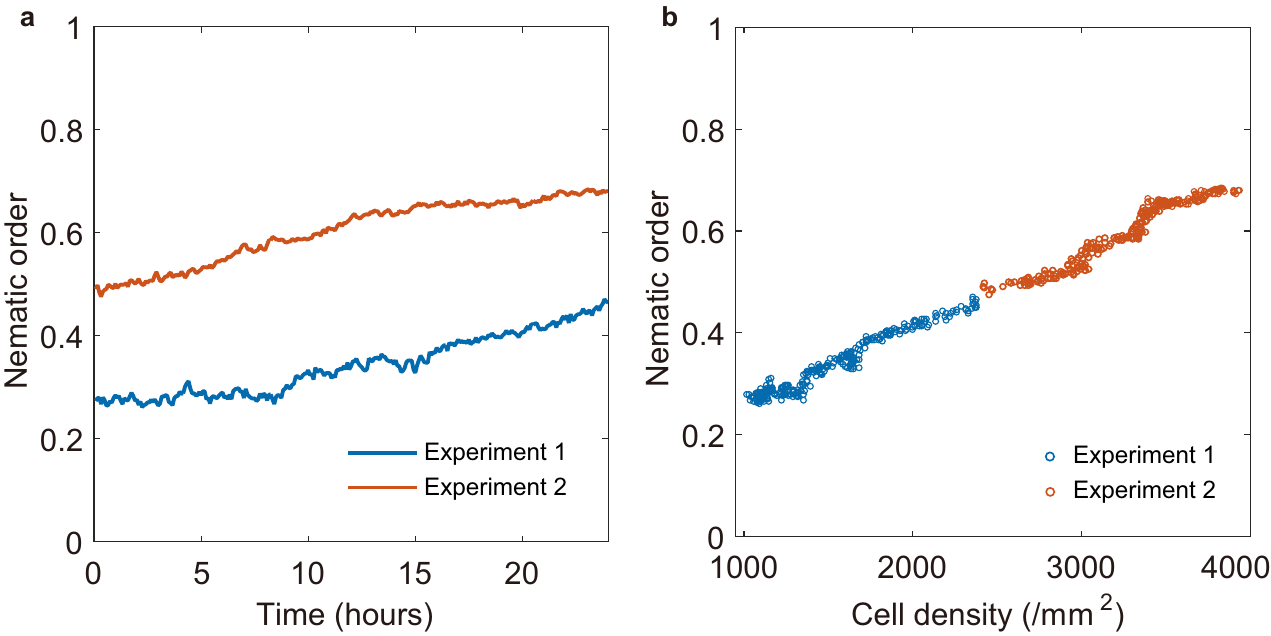}
 \end{center}
  \caption{\label{FigS1}
{\bf Growing nematic order in NPC culture.}
{\bf a,} Nematic order [see Eq.~(\ref{nematicorder})] calculated for a fixed system size $L=185\ \mu{\rm m}$ (chosen as an intermediate size that is larger than a single cell and smaller than the typical distance between defects) for each time frame in the two separate experiments.
{\bf b,} Cell density vs nematic order.
}
\end{figure*}

\begin{figure*}[!t]
 \begin{center}
  \includegraphics[width=100mm]{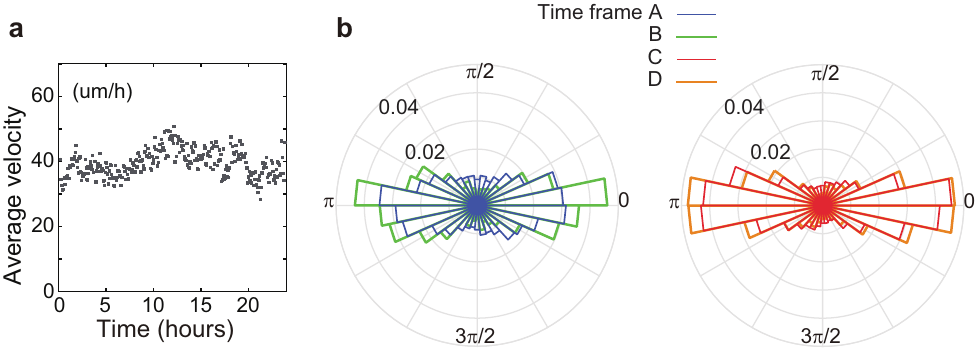}
 \end{center}
  \caption{\label{FigS2}
{\bf Velocity and angle of cell motion.}
{\bf a,} Time-dependence of average cell velocity calculated from the displacement of cells between frames from experiment 2 (high cell density, time frames C and D shown in Fig.~2a). {\bf b,}  Distribution of direction of motion of cells with respect to the longitudinal direction calculated for each cell trajectory. See Supplementary Text for detail.}
\end{figure*}

\begin{figure*}[!t]
 \begin{center}
  \includegraphics[width=160mm]{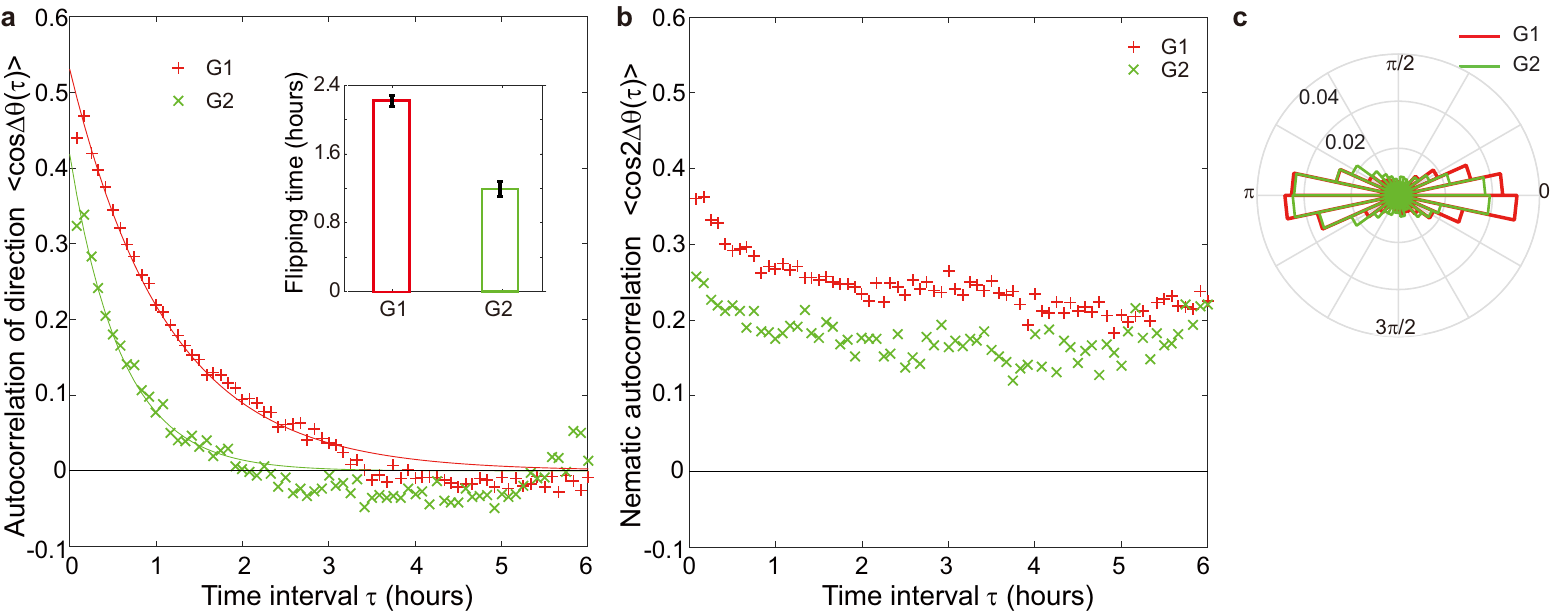}
 \end{center}
  \caption{\label{FigS3}
{\bf Velocity reversal during the same cell cycle phase.}
{\bf a,} Autocorrelations, {\bf b,} nematic autocorrelations, and {\bf c,} distribution of velocity angle relative to the longitudinal angle calculated for cells in different cell cycle phase (G1 and G2) from the same experiment of Fucci-labeled NPC at high cell density ($>$3000 cells mm$^2$). Inset in {\bf a} is the flipping time quantified from the fitting of autocorrelation by exponential. Error bars correspond to the standard error of fitting. See also Supplementary Movie 3.}
\end{figure*}

\begin{figure*}[!t]
 \begin{center}
  \includegraphics[width=130mm]{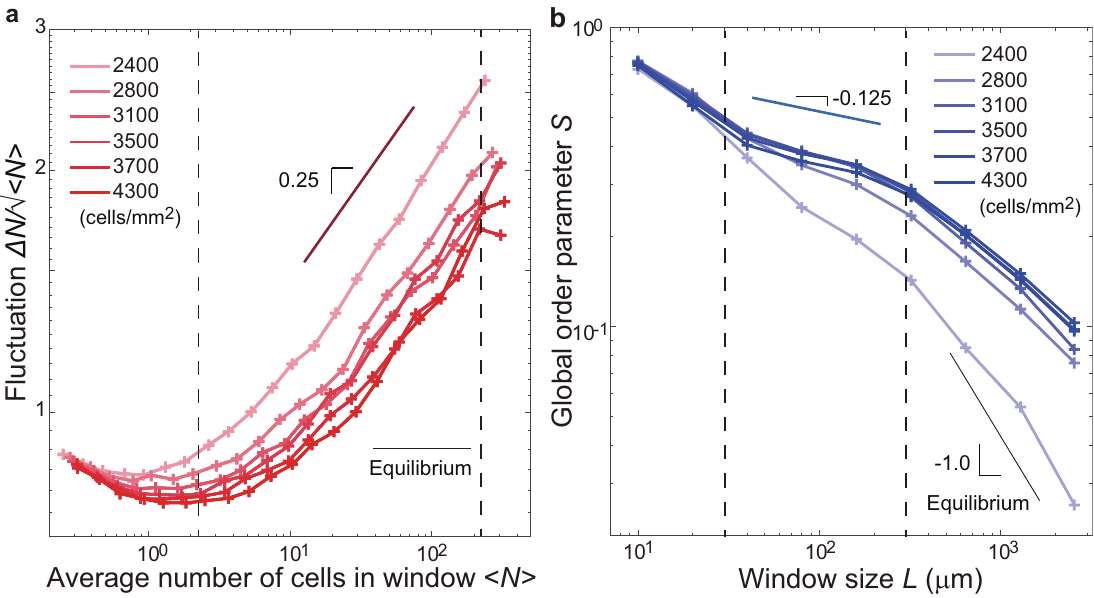}
 \end{center}
  \caption{\label{FigS4}
{\bf Giant number fluctuation and global order parameter.}
{\bf a,}  Giant cell number fluctuation in the dense culture (time frames C and D in Fig.~2a). The fluctuation rescaled by the square root of the mean cell number in the different defined system sizes were calculated from counting of the cell nuclei.
{\bf b,} Global nematic order as a function of system size [see Eq.~(\ref{nematicorder})]  in the dense culture. Nematic order was calculated from the velocity field obtained by the global tracking algorithm. 
In both figures, the range between dotted lines indicate the system size larger than the typical cell size and smaller than the typical distance between defects ($L \sim$30-300$\mu$m).
}
\end{figure*}

\begin{figure*}[!t]
 \begin{center}
  \includegraphics[width=150mm]{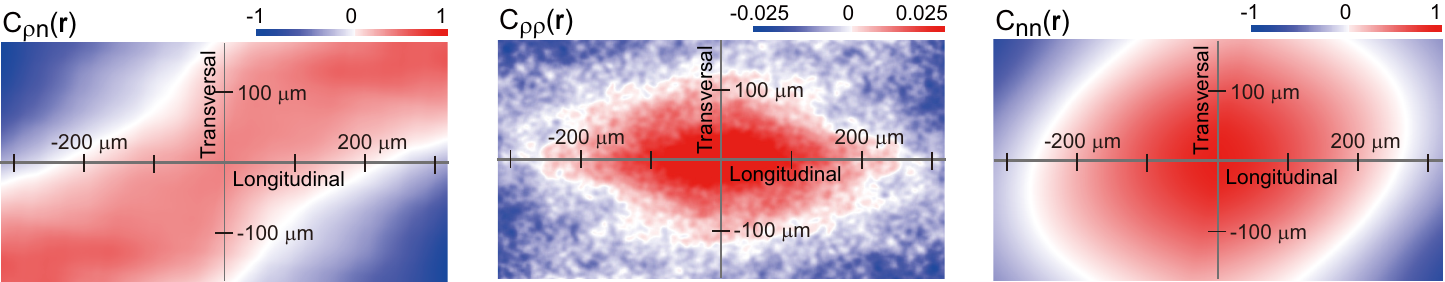}
 \end{center}
  \caption{\label{FigS5}
{\bf Spatial correlation of the fluctuation of density and alignment.}
Spatial correlations of density fluctuation and alignment fluctuation calculated in the ordered region without defect (size $\sim$0.2~mm$^2$, cropped from images in time frames C and D in Fig.~2a). The values are normalized using the standard deviations of the density and alignment fluctuations [see Eqs.~(\ref{eq:corrrhorho}-\ref{eq:corrnn}) for definition]. The horizontal axis corresponds to the alignment direction defined in the analyzed regions. Average of 4 regions. 
}
\end{figure*}

\begin{figure*}[!t]
 \begin{center}
  \includegraphics[width=181.45mm]{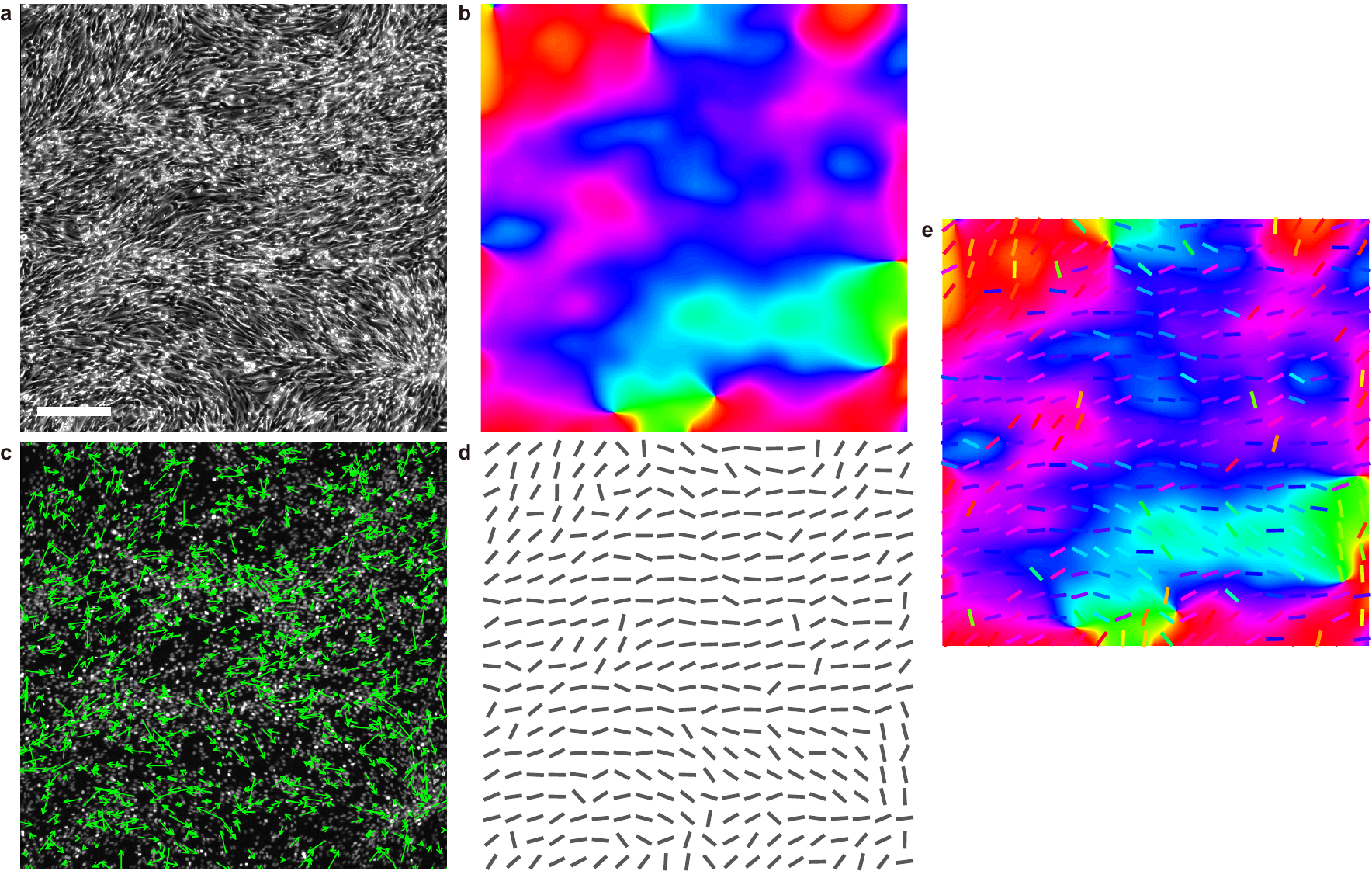}
 \end{center}
  \caption{\label{FigS6}
{{\bf Comparing the tensor method and the tracking method to extract alignment from image.} {\bf a,} Phase contrast image of NS culture in the high density regime. Scale bar = 200 $\mu$m. {\bf b,} Direction of alignment obtained by applying the tensor method on the phase contrast image data. {\bf c,} Fluorescent image of the cell nucleus (white) in the same position as {\bf a}, overlaid with the displacements (green arrows) in a single time frame (5 mins interval).  Only showing $\sim 20\%$ of the calculated displacements (elongated 15-fold) for the sake of visualization. Scale bar = 200 $\mu$m. {\bf d,} Alignment field obtained from the nematic order calcuated within the boxes (60 pixels$\times60$ pixels = $54.7 \times 54.7~\mu {\rm m}^2$) using the displacement vectors. {\bf e,} Merge of the alignment fields obtained by two independent methods. Bars indicating the direction of alignment (same as {\bf d}) were colored according to the same rule as {\bf b} to see match between the two fields.}
} 
\end{figure*}

\begin{figure*}[!t]
 \begin{center}
  \includegraphics[width=120mm]{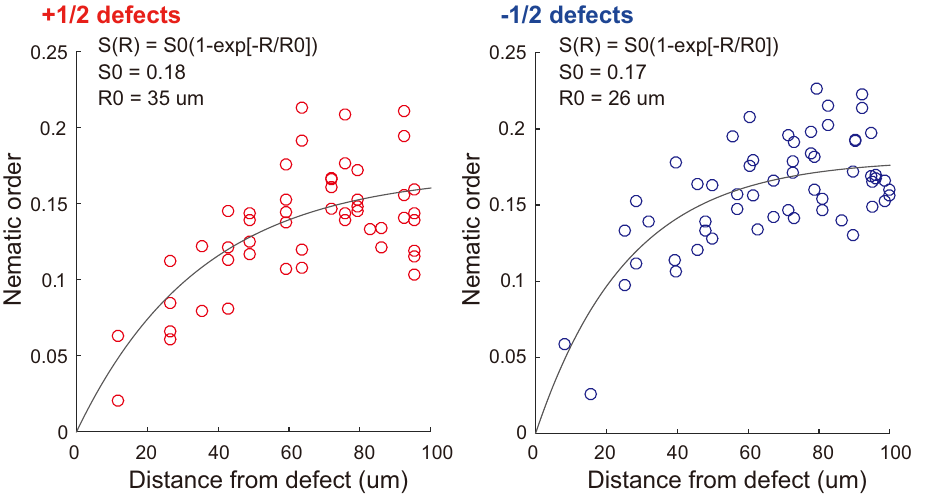}
 \end{center}
  \caption{\label{FigS7}
{\bf Nematic order around defects.}
Fit of the nematic order near the defect calculated from the data points obtained by the tracking method. A radially symmetric form of the nematic order is assumed (see Supplementary Text).
}
\end{figure*}

\begin{figure*}[!t]
 \begin{center}
  \includegraphics[width=120mm]{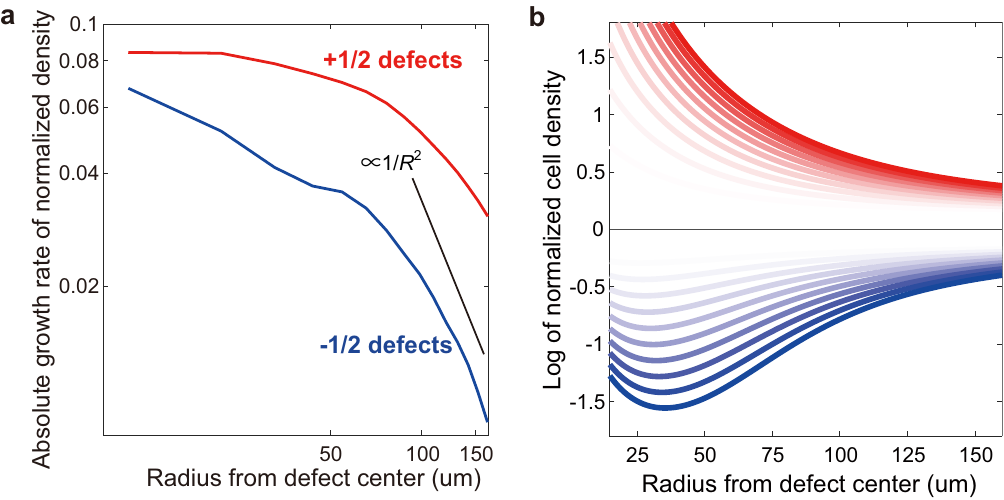}
 \end{center}
  \caption{\label{FigS8}
{\bf Growth rates around defects compared with theory.}
 {\bf a,} Log-log plot of the absolute growth rates around the defects  (same as Fig.~4c). {\bf b,} Time evolution of density around the defects obtained by simulation with parameters of the defects obtained from fitting (Fig.~S6). We used $\Delta/\gamma_0 = 0.2$ and set $t$ in a.u. (but with equal intervals) to see roughly the same growth regime as obtained in experiment (Fig.~4b).
}
\end{figure*}

\begin{figure*}[!t]
 \begin{center}
  \includegraphics[width=100mm]{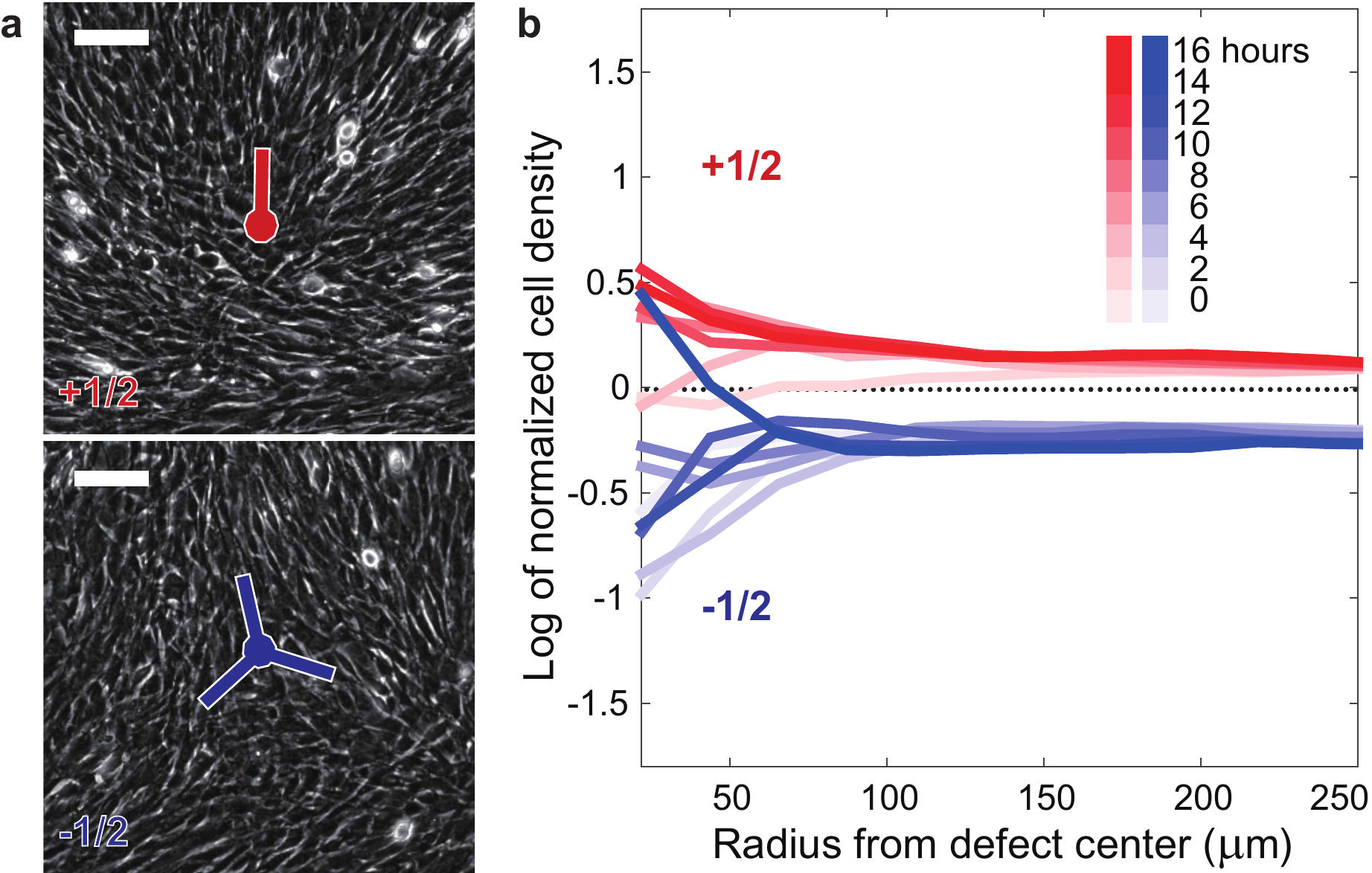}
 \end{center}
  \caption{\label{FigS10}
{\bf Defect  dynamics of myoblast cells (C2C12).}
{\bf a,} Phase contrast image of C2C12 cells, highlighting the existence of +1/2 and -1/2 topological defects in the ordered pattern. {\bf b,} Cell density dynamics around the defects. Average of 4 defects each from the same dish over the same time course.
}
\end{figure*}
\end{document}